\documentclass{ws-mplb}

\usepackage{cancel}

\newcommand{\bn}{\begin{enumerate}}
\newcommand{\en}{\end{enumerate}}
\newcommand{\ba}{\begin{eqnarray}}
\newcommand{\ea}{\end{eqnarray}}

\usepackage{graphicx,color}

\newcommand{\be}{\begin{equation}}
\newcommand{\ee}{\end{equation}}
\newcommand{\la}{\langle}

\newcommand{\ra}{\rangle}

\newcommand{\ete}{{\it et al.}}

\begin{document}

\markboth{Zhang, Bai, and George }{
Spin Berry points
}

%
\catchline{}{}{}{}{}
%

\title{ Spin Berry points as crucial for ultrafast
  demagnetization
}

\author{G. P. Zhang$^*$
}

\address{Department of Physics, Indiana State University,
   Terre Haute, IN 47809, USA
$^*$guo-ping.zhang@outlook.com}

\author{Y. H. Bai}

\address{Office of Information Technology, Indiana State
  University, Terre Haute, IN 47809, USA}

\author{Thomas F. George}

\address{Departments of Chemistry \& Biochemistry and Physics \&
  Astronomy, University of Missouri-St. Louis, St. Louis, MO 63121, USA}

\maketitle

\begin{history}
\received{(\today)}
\revised{(Day Month Year)}
\end{history}

\begin{abstract}
{ Laser-induced ultrafast demagnetization has puzzled researchers
  around the world for over two decades.  Intrinsic complexity in
  electronic, magnetic, and phononic subsystems is difficult to
  understand microscopically. So far it is not possible to explain
  demagnetization using a single mechanism, which suggests a crucial
  piece of information still missing. In this paper, we return to a
  fundamental aspect of physics: spin and its change within each band
  in the entire Brillouin zone. We employ fcc Ni as an example and use
  an extremely dense {\bf k} mesh to map out spin changes for every
  band close to the Fermi level along all the high symmetry lines. To
  our surprise, spin angular momentum  at some special {\bf
    k} points abruptly changes from $\pm \hbar/2$ to $\mp \hbar/2$
  simply by moving from one crystal momentum point to the next. This explains why intraband transitions, which the spin
  superdiffusion model is based upon, can induce a sharp spin moment
  reduction, and why electric current can change spin orientation in
  spintronics.  These special {\bf k} points, which are called spin
  Berry points, are not random and appear when several bands are
  close to each other, so the Berry potential of spin majority states
  is different from that of spin minority states. Although within a
  single band, spin Berry points jump, when we group several
  neighboring bands together, they form distinctive smooth spin Berry
  lines.  It is the band structure that disrupts those lines. Spin
  Berry points are crucial to laser-induced ultrafast demagnetization
  and spintronics.}
\end{abstract}


\section{Introduction}

Laser-induced ultrafast demagnetization on the femtosecond time
scale\cite{eric} is interesting because it challenges the current
wisdom about the physics of magnetism and ultrafast laser excitation,
and because it presents a new opportunity to manipulate spin on an
unprecedented time scale. Various mechanisms have been
proposed\cite{prl00,koopmans2010,battiato2010,dornes2019}, with no
agreement among these mechanisms. In the literature, there are at
least three pictures, with many questions remaining unanswered.  The
first is the local picture where the electron spin is quenched on the
site. There are several ways as to how such quenching occurs: (i) The
electron may pass the angular momentum to the lattice within several
hundred femtoseconds.\cite{dornes2019} However, it is well known that
phonons only participate in the dynamics on a picosecond time scale,
not femtosecond. If phonons are the key to demagnetization, one should
expect this fast angular momentum transfer in half-metallic
CrO$_2$\cite{zhang2006} and Heusler compounds.\cite{muller2009}
Instead, experimentally it is found that the demagnetization time is
related to spin polarization. It is likely that this phonon mechanism
is only part of the story.\cite{prb08,prb09,stamm2010} (ii) The
electron may flip spins through spin-orbit
coupling,\cite{prl00,koopmans2010} but the spin-orbit coupling may
flip spins in both directions ($L^+S^-+L^-S^+$), i.e.,
bidirectional. So there is no guarantee that spin must be
reduced. (iii) The electron may transfer the spin angular momentum to
orbital.\cite{prl00,tows2015,dewhurst2020b} This was our original
idea,\cite{prl00} which was further explored in different settings
lately.\cite{tows2015,dewhurst2020b} This mechanism has no issue with
the time scale, but is difficult to prove experimentally. Orbital
momentum change mainly occurs when the laser field is on.\cite{jap19}
Once the laser field is over, the system evolves within the relative
subspaces of spin and orbital. The second difficulty is
experimental. In time-resolved x-ray magnetic circular dichroism,
there is no experiment that ever reported an increase in orbital
momentum,\cite{boeglin2010,bergeard2014} to the best of our
knowledge. On the other hand, the absence of orbital angular momentum
increase should not be used as evidence of no angular momentum
transfer, because the total angular momentum ${\bf J=L+S}$ is not a
conserved quantity in solids. A reduction in {\bf S} does not
necessarily lead to an increase in {\bf L}, because they both can be
reduced, but just not at the same rate.

The second is the nonlocal picture where the electron spin is moved
out of the region that is excited. This is an itinerant picture where
electrons are mobile.\cite{jpcm18} The phenomenological spin
superdiffusion model\cite{battiato2010} is based on this picture. The
majority spin moves faster than the minority spin, so the deficiency
of the majority spin after laser excitation leads to demagnetization.
Because the model is phenomenological, the source term that starts the
entire demagnetization is lumped into a single term,\cite{battiato2012}
where no actual laser pulse is present in the theory.

One limitation of the above mechanisms is that they require
significant energy absorption. However, experimentally a significant
spin reduction is already found\cite{jpcm10} even with fluence of
around 1 mJ/cm$^2$.  This energy constraint is important as it puts a
limit on a theory, so that the incident laser fluence is not used as a
fitting parameter.  This brings us to the third picture: the spin wave
or magnon picture.\cite{turgut2016,jap19} Here spins are situated at a
lattice site and do not move, but their orientations can be changed
in space.  This directly generates a long-wavelength spin waves across
many hundred lattice sites. We recently found that this
demagnetization is extremely efficient.\cite{jap19} To this end, three
existing pictures may not be enough to cover all the possible channels to
demagnetization. It is imperative to explore new channels.

In this paper, we go back to the basic physics in a
ferromagnet. Regardless of what the true underlying mechanism is, the
basic spin properties remain the same. We take fcc Ni as an example
and examine how spin angular momentum changes in those bands around
the Fermi level and across the entire Brillouin zone. To our surprise,
by simply moving from one crystal momentum {\bf k}
point to another within the same band, spin angular momentum undergoes a radical jump from
$+\hbar/2$ to $-\hbar/2$, or from $-\hbar/2$ to $+\hbar/2$. These {\bf
  k} points are not random, and we call them the spin Berry
points because they are with the band curvature change.
When we group them from neighboring bands, they form smooth lines, namely
spin Berry lines. Each high symmetry direction has its unique set of spin
Berry points. Along the $\Gamma$-X direction, we have four such spin
Berry points, from $\alpha$ to $\gamma$. These Berry points delineate
where spins can be flipped through intraband or interband
transitions. Microscopically, in presence of spin-orbit coupling, the
spin operator does not permute with the total Hamiltonian. The band
structure separates spin sectors, so spin angular momentum from the
same spin sector appears at two or more different bands. These spin
discontinuities are important since they allow the intraband transition
to change spin easily. This is the same reason why electric current
can change spin orientation in spintronics.

The rest of the paper is arranged as follows. Section 2 is devoted to
the interaction between the system and laser field, followed by our
theoretical formalism in Sec. 3. Our main results are in Sec. 4, where
the band structure and spin angular momentum change along high symmetry
directions, and spin Berry points are given. We discuss our results in
Sec. 5. Finally, we conclude this paper in Sec. 6.




\newcommand{\rr}{{\bf r}}

\newcommand{\ik}{i{\bf k}}
\newcommand{\jk}{j{\bf k}}

\newcommand{\nk}{\psi_{n{\bf k}_2}(\rr)}
\newcommand{\mk}{\psi_{m{\bf k}_1}(\rr)}

\section{Interaction with the laser field}


Laser pulses are an electromagnetic wave.  Since it is a transverse
wave, the propagation direction is perpendicular to the electric field
{\bf E} and magnetic field {\bf B}.  To describe the laser interaction
with the system, we change the momentum ${\bf p}$ to ${\bf p} +e{\bf
  A} (\rr,t)$, where ${\bf A} (\rr,t)$ is the laser vector
potential. This way the interaction Hamiltonian $H_I$ is \be
H_I=\frac{e}{m_e}\left ({\bf p}\cdot {\bf A} + {\bf A}\cdot {\bf
  p}\right ) + \frac{e^2A^2}{2m_e}. \ee To be definitive, we choose
${\bf A} (\rr,t)={\bf A}_0(t) {\rm e}^{i({\bf q}\cdot\rr-\omega t)}, $
where {\bf q} is the light photon wavevector. This is the classical
form of the vector potential. In the second quantization, the vector
potential is replaced by two photon operators ($a^\dagger,a$) as
$\hat{\bf A}={\bf A}_0(t)(a^\dagger {\rm e}^{i({\bf q}\cdot\rr-\omega
  t)} + a {\rm e}^{-i({\bf q}\cdot\rr-\omega t)})$\cite{scully}, but
the spatial and temporal dependences ${\rm e}^{i({\bf
    q}\cdot\rr-\omega t)}$ remain unchanged. It must be stated clearly
that once {\bf q} is chosen, both the electric field and magnetic
field directions must be perpendicular to {\bf q}.

We consider the transition matrix elements of ${\bf p}\cdot {\bf
  A}(\rr,t)$ between two band states $\mk$ and $\nk$, \ba &&\int_{\rm
  all ~space} d\tau \mk^*{\bf p}\cdot {\bf A}(\rr,t)
\nk  =\sum_{{\bf R}_l} {e}^{i(-{\bf k}_1+{\bf q}+{\bf k}_2)\cdot
  {\bf R}_l} \nonumber\\ &\times& \int_{\rm u.~c.}d\tau u_{n{\bf k}_1}(\rr)^* {e}^{-i{\bf
    k}_1\cdot \rr} {\bf p}\cdot {\bf A}_0(t) {e}^{i({\bf q}\cdot
  \rr-\omega t)} u_{m{\bf k}_2}(\rr) {e}^{i{\bf k}_2\cdot \rr}, \ea
where $u_{n{\bf k}}$ is the periodic part of the Bloch wavefunction
${\psi_{n{\bf k}}(\rr)}$
and the second integration is over the unit cell only.
The summation over the entire lattice site ${\bf R}_l$ gives
$\delta_{-{\bf k}_1+{\bf q}+{\bf k}_2,0}$, which guarantees the
conservation of the momentum in terms of the crystal momentum.  The
crystal momentum change is the photon momentum.  We make two comments
on this. (i) This conservation is restricted to the wave nature of the
electron. Since the crystal momentum in solids is not the same as the
momentum of the electron, we only can say that there is a small
crystal momentum shift during the transition from one band to
another. For a laser pulse of wavelength $\lambda$, $|{\bf
  q}|=2\pi/\lambda$ should be compared with the reciprocal wavevector
${\bf b}$ in the Brillouin zone. In a fcc structure, $|{\bf
  b}|=2\pi/|{\bf a}|$, where $a=|{\bf a}|$ is the lattice constant,
not to be confused withe creation and annihilation operators
above. Suppose $\lambda=800$ nm and $a=3.524~\rm\AA$ (fcc Ni's lattice
constant). Then $|{\bf q}/{\bf b}|=a/\lambda=0.0004405$, which is
extremely tiny, so normally one has ${\bf k}_{1}\approx {\bf
  k}_2$. This favors the vertical interband excitation but ignores
intraband transitions in metals. Intraband transitions rely on this
tiny change in the crystal momentum in the band states to move from
one {\bf k} point to another.  For instance, if a laser pulse
propagates along the $-z$ axis (see Fig. \ref{fig1}(a)), {\bf q} is
$-q_z$, so that the crystal momentum changes along the $k_z$
direction.

Another change is in the mechanical momentum ${\bf p}+e{\bf A}$, where
{\bf p} is the canonical momentum and ${\bf A}$ is the laser field vector potential. This is how we obtain
the interaction Hamiltonian with the laser field above.  The vector
potential also enters intraband transitions through {\bf
  k}.\cite{jpcm18} If we ignore the spatial dependence of the vector
potential and consider a single band, the crystal momentum ${\bf k}$
becomes time-dependent and can be approximately written
as,\cite{jpcm18} \be {\bf k}(t)+e {\bf A}(t)/\hbar={\bf
  k}_0, \label{k} \ee where ${\bf k}_0$ is the initial crystal
momentum.  Since ${\bf A}$ is perpendicular to {\bf q}, here the
crystal momentum direction is different from the above. If light
propagates along the $-z$ axis, only $k_x$ and $k_y$ are changed
(see. Fig. \ref{fig1}(a)). We should point out that Eq. \ref{k} is not
generic. This new ${\bf k}(t)$ is seemingly similar to ${\bf p}+e{\bf
  A}$ divided by $\hbar$, which might prompt one to rewrite the band
energy $E_n$ as $E_n[({\bf p}+e {\bf A})/\hbar]$, which is not
strictly correct.\cite{callaway} Even if our external laser field is
zero (both {\bf E} and {\bf B} are zero), {\bf A} is not necessarily
zero and can be a nonzero constant both in space and time.  However,
if we stick with a single band $E_n({\bf k}(t))$, the energy change
with time \be \frac{\partial E_n}{\partial t}
=-\frac{e}{\hbar}\nabla_k E_n \cdot \frac{\partial {\bf A}}{\partial
  t} =\frac{e}{m_e} {\bf p}_{nn} \cdot {\bf E}, \label{energy} \ee
where we use ${\bf v}_{nn}=\nabla_k E_n/\hbar={\bf p}_{nn}/m_e$. This
shows that the band energy changes with the electric field and this
change is gauge-invariant, because the band energy is measurable.

Equation \ref{energy} is quite insightful. This leads
us to investigate how spins are changed under the same laser field
excitation.

\section{Theoretical formalism}

We employ the state-of-the-art density functional theory and solve the
Kohn-Sham equation,\cite{wien2k,blaha2020,np09,prb09} \be \left
[-\frac{\hbar^2\nabla^2}{2m_e}+V_{ne}+V_{ee}+V_{xc} \right
]\psi_{\ik}(\rr)=E_{\ik} \psi_{\ik} (\rr), \label{ks} \ee where $m_e$
is the electron mass, and the terms on the left side are the kinetic
energy, nuclear-electron attraction, electron-electron Coulomb
repulsion and exchange correlation, respectively.  We include the
spin-orbit coupling through the second-variational principle.  After
we include the spin-orbit coupling, all the states are
spin-mixed. Here $\psi_{\ik}(\rr)$ is the Bloch wavefunction of band
$i$ at crystal momentum ${\bf k}$, and $E_{\ik}$ is the band energy.
We utilize the full-potential augmented plane wave method as
implemented in the Wien2k code,\cite{wien2k} where within the
Muffin-tin sphere atomic basis functions are used, and in the
interstitial region a plane wave is used. This method is among the
most accurate methods available.

We can define a similar spin change as Eq. \ref{energy}, \be
\frac{\partial s^z_{n{\bf k}}}{\partial t}=\nabla_k s^z_{n{\bf k}}
\cdot \frac{e}{\hbar} \frac{\partial {\bf A}(t)}{\partial t}
=-\frac{e}{\hbar} \nabla_k s^z_{n{\bf k}} \cdot {\bf E}(t).
\label{spin}\ee
The unit of spin angular momentum change rate is Nm, a torque that is
delivered by an electric field.  This equation shows that the spin
change rate follows the electric field of the laser pulse, but whether
one has a spin reduction or enhancement depends on the spin dispersion
with crystal momentum {\bf k}.  Here $s^z_{n{\bf k}}$ is computed as
$\la \psi_{n{\bf k}} |\hat{s}_z| \psi_{n{\bf k}}\ra$, for band state
$\psi_{n{\bf k}}$.  In the following, we investigate this spin
dispersion.

\section{Results}

\subsection{Band structure close to the Fermi level}

\begin{figure}
\centerline{\psfig{file=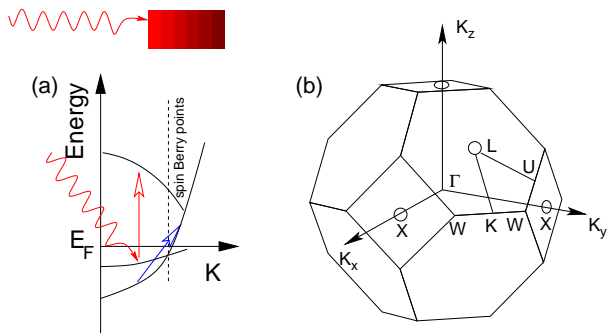,width=8cm,angle=0}}
\caption{(a) Two different laser-induced excitations. One is an intraband
  transition, where electrons only move within the same band. The
  other is an interband transition between two different bands. What
  is important to demagnetization is whether these transitions alter
  spin characters. Top: laser incident on a sample.
(b) Three-dimensional Brillouin zone in a fcc
  lattice. The laser propagates along the $-k_z$ axis, with {\bf E}
  and {\bf B} in the $k_x$-$k_y$ plane.  The spin Berry lines are
  introduced here to understand spin changes during transitions.  }
\label{fig1}
\end{figure}

We take fcc Ni as an example.  The product of the Muffin-tin radius
and plane wave cutoff is 9.5, and we include one local orbital at 0.69
Ryd.  As stated above, we include spin-orbit coupling (SOC) at the
second variational principle level. This means that we use the
spin-polarized band states $\phi_{i{\bf k},\sigma}(\rr)$, where
$\sigma$ denotes the spin orientation, as the basis for the SOC
calculation. The final wavefunction is expressed as $\psi_{n{\bf
    k}}(\rr) =\sum_{i,\sigma} c_{i{\bf k},\sigma}^n \phi_{i{\bf
    k},\sigma}(\rr)$.  This has a benefit that we can extract the spin
matrix elements of spin operators ($S_x$, $S_y$ and $S_z$) between
bands from $c_{i,\sigma}^n$. We express $S_x$, $S_y$ and $S_z$ in
terms of Pauli matrices, $\sigma_x$, $\sigma_y$ and $\sigma_z$. The
{\bf k} points along high symmetry lines contain coordinates at 1/4
the reciprocal lattice vector {\bf b} (see below), so our number of
{\bf k} points along $k_x$, $k_y$ and $k_z$ must be a multiple of 4.
A dense {\bf k} mesh grid is necessary.  In this work, we employ a
mesh of $104\times 104 \times 104 $. There are 73763 irreducible
points.  There are eight special symmetry lines. Figure \ref{fig1}(b)
shows the entire Brillouin zone with the $\Gamma$ point at
(0,0,0). All the following coordinates are expressed in the
conventional reciprocal lattice vector of a simple cubic, not in a fcc
primitive lattice.  The length of the reciprocal lattice vector in the
conventional lattice is twice that of the primitive lattice.

The X point is at (1,0,0). There are in total six X
points at ($\pm$1,0,0), (0,$\pm$1,0), and (0,0,$\pm$1), but because
(1,0,0) and (-1,0,0) differ by one reciprocal lattice vector, they count
as one point. The net number of X points reduces from 6 to 3.

The W points are at (1,$\frac{1}{2}$,0), with the distance from the
$\Gamma$ point being $(\sqrt{5}/2) 2\pi/a$. They are at the six
corners of the hexagons. There are eight hexagons, but two hexagons
share two W points, so there are $6\times 8/2=24$ W points in
total. These eight hexagons are enough to create the entire
Brillouin. At the middle of two neighboring W points are K points. For
two W points at (1/2,1,0) and (1,1/2,0), K is at (3/4,3/4,0). U is at
the middle of another two W points (1/2,1,0) and (0,1,1/2), so its
coordinate is (1/4,1,1/4). U and K have the same distance to the
$\Gamma$ point and are approximately equivalent.

\begin{figure}
\centerline{\psfig{file=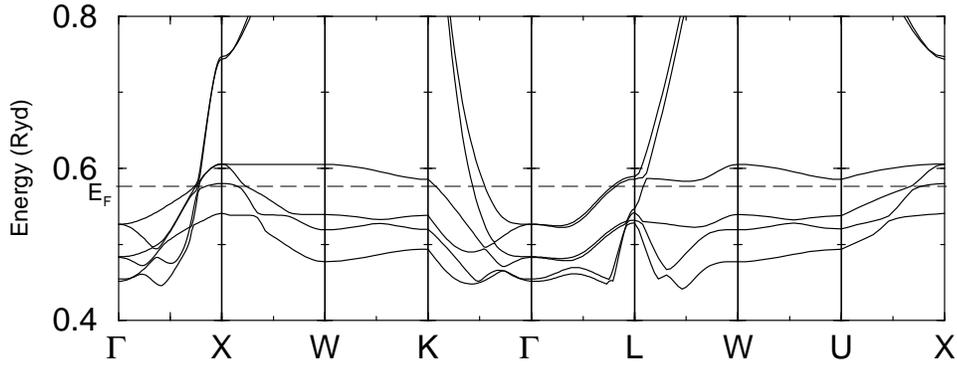,width=1\columnwidth,angle=0}}
\caption{
Band structure close to the Fermi level. The horizontal dashed  line
denotes the Fermi energy at 0.57642 Ryd.
 }
\label{bandstructure}
\end{figure}

Figure \ref{bandstructure} shows the band structure close to the Fermi
level (see the horizontal dashed line), with $E_F$ at 0.57648 Ryd,
which is fully consistent with our prior calculation.\cite{prb09} One
can see the band structure along the W-K and W-U lines is similar.
Their bands are away from the Fermi level.  Ni has 28 electrons, where
10 electrons are in the core shell, and the remaining 18 electrons are
in the valence bands.  10 valence electrons fill the topmost $3d$ and
$4sp$ bands.  Since we are interested in intraband transitions, we
only show band states close to the Fermi level.

An important feature among these bands is that they cross each other
frequently. For instance, along the $\Gamma$-X direction, $4sp$ bands,
which have a large dispersion, cross the $3d$ bands which are less
dispersive. This crossing has an important consequence on the spin
angular momentum.  The Fermi level cuts through bands along the
$\Gamma$-X, X-W, K-$\Gamma$, $\Gamma$-L, L-W, and U-X directions. This
provides an avenue for intraband transitions, where the electron can
move within bands. Investigating how the spin character changes as one
moves from one {\bf k} point to another is valuable, since it gives us
the first hint how laser-driven charge transport leads to
magnetization change, as illustrated in Fig. \ref{fig1}(a).

\subsection{Spin change along the W-K and W-U directions}

While the band dispersion in fcc Ni is well known, little is known as to how
each band state changes its spin orientation as we straddle through
the Brillouin zone. King \ete\cite{king2014} show that in SrTiO$_3$,
there is a special spin-orbital texture around the Fermi surface, but
SrTiO$_3$ is an insulator, so there are no bands across the Fermi
level. For fcc Ni, nothing is known at present.  In particular, the
Berry phase may change suddenly, so the spin change may show a spike.
We decide to choose a simple case where bands in the Brillouin zone do
not cross the Fermi level.  Naturally, they do not effectively
contribute to the spin moment change, but provide a starting point to
understand the underlying physics.

\begin{figure}
\centerline{\psfig{file=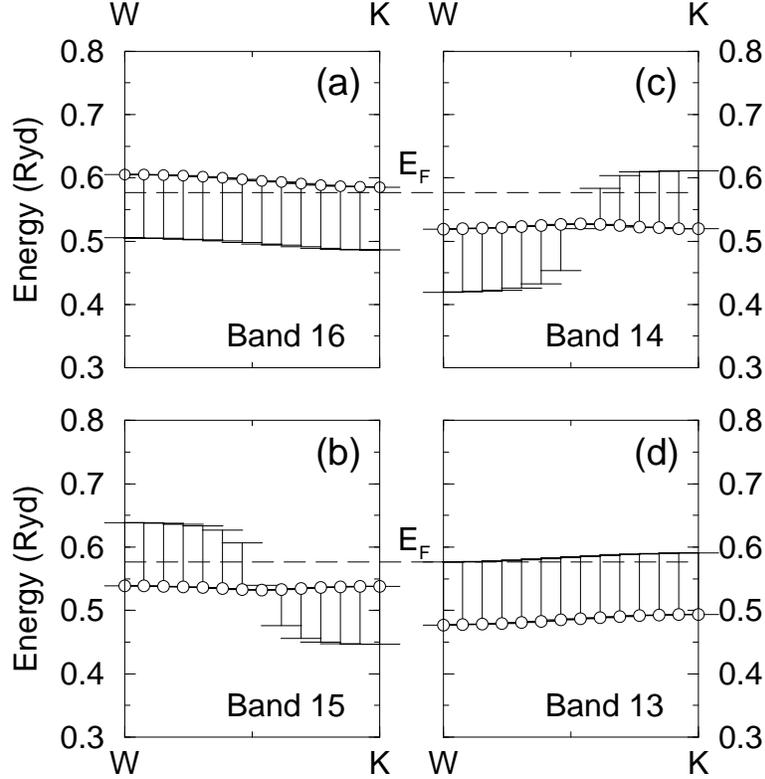,width=0.8\columnwidth,angle=0}}
\caption{Spin angular momentum along the W-K direction for four bands
  that are close to the Fermi surface denoted by horizontal dashed
  lines.  The circles denote the band energy. The error bar denotes
  the amplitude of the spin angular momentum at each band energy for
  every band. The longest bar corresponds to $\hbar/2$.  If the bar
  is below circles, this means that the spin value is negative; if
  above, positive.  Our interest is to see whether spins start to
  flip.  (a) Band 16 and (d) band 13 have a weak spin change from W to
  K since these two bands are away from other bands with little spin
  mixing.  (b) Band 15 and (c) band 14 appear in the same energy
  window, so they have a strong change in spin angular momentum.}
\label{w2k}
\end{figure}

 As discussed above, band dispersions are similar along the W-K and
 W-U directions. We superimpose the spin angular momentum on the bands
 which are denoted by the empty circles (Fig. \ref{w2k}). The error bars do not
 represent the error in the band energy; instead they represent spin
 angular momentum. If the bar is above a circle, this means that the
 spin angular momentum is positive; below a circle, negative. The
 length of the bar is proportional to the magnitude of spin
 $s^z_{n{\bf k}}$, with the longest bar corresponding to
 $\hbar/2$. Figure \ref{w2k}(a) shows that band 16 has all negative
 spins, while band 13 (Fig. \ref{w2k}(d)) has positive spins from W to
 K. This is understandable. Since they are away from other
 bands, with or without SOC, there is little interaction to change
 their spin characters.  Because energetically majority and minority
 bands are shifted with respect to each other due to the exchange
 interaction, they have different occupations.

\begin{figure}
\centerline{\psfig{file=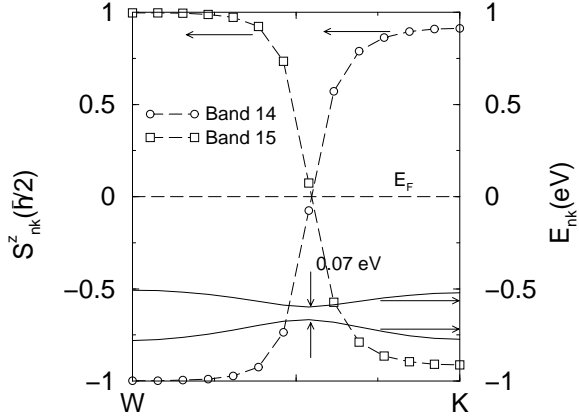,width=0.6\columnwidth,angle=0}}
\caption{Spin angular momentum $s^z_{n{\bf k}}$ in bands 14 and 15
  along the W-K direction (use the left scale).  The circles refer to
  spins for band 14 and the boxes for band 15. The Fermi level in this
  figure is set at zero (horizontal dashed line), and the band energy
  (two solids lines) is in unit of eV (right scale).  }
\label{w2ka}
\end{figure}

By contrast, Figs. \ref{w2k}(b) and \ref{w2k}(c) show that bands 14
and 15 are in the same energy window, so we expect there is a strong
interaction between them.  Figures \ref{w2k}(b) and \ref{w2k}(c) show
that their $s^z_{n{\bf k}}$ changes sign going from W to K; and in the
middle of the zone, both bands have a very small spin, (see how small
those bars are in Figs. \ref{w2k}(b) and \ref{w2k}(c)).  We plot the
spin change for these two bands in Fig. \ref{w2ka}. It is obvious that
$s^z_{n{\bf k}}$ for these two bands is symmetric with respect to 0
(compare boxes with circles), and $s^z_{14,{\bf k}}=-s^z_{15,{\bf k}}$
along the W-K line.  $s^z_{14,{\bf k}}$ can be fitted to
$\tanh[(-k-k_0)/c]$, where $k_0$ and $c$ are two constants. This
indicates that the orbital contribution from these two bands is
relatively simple, so the spin-flipping term in spin-orbit coupling
${\bf L}\cdot {\bf S}$ plays a decisive role through \ba
|\psi_1\ra&=&|\phi^\uparrow \ra + \lambda\frac{\la \phi^\downarrow
  |{\bf L}\cdot {\bf S}|\phi^\uparrow \ra }{\Delta E}|\phi^\downarrow
\ra =|\phi^\uparrow \ra + \lambda\overbrace{\frac{ \la \phi^\downarrow
    | L^+ S^-|\phi^\uparrow \ra }{\Delta E}}^{spin-flipping~down}
|\phi^\downarrow \ra \\ |\psi_2\ra&=&|\phi^\downarrow \ra + \lambda
\frac{\la \phi^\uparrow |{\bf L}\cdot {\bf S}|\phi^\downarrow \ra
}{\Delta E}|\phi^\uparrow \ra =|\phi^\downarrow \ra + \lambda
\overbrace{\frac{\la \phi^\uparrow |L^- S^+|\phi^\downarrow \ra
  }{\Delta E}}^{spin-flipping~up}|\phi^\uparrow \ra \ea where
$\lambda$ is the spin-orbit coupling, $|\phi^\uparrow\ra$ and
$|\phi^\downarrow\ra$ are the spin-polarized wavefunctions, and
$\Delta E$ is the gap between spin majority and minority bands.  Two
flipping terms depend on {\bf k}. It should be noted that in this
case, angular momentum transfer from spin to orbital is revealed
through operators $L^{\pm}$. It is equally important to point out that
one should not expect that the spin and orbital angular momentum
changes completely compensate each other, because in solids spherical
symmetry is broken.\cite{prb08,prb09} The minimum gap between these
two bands is 0.07 eV, which matches the spin-orbit coupling constant
of 0.07 eV in our prior model calculation.\cite{prl00} These bands are
occupied, but if excitation leads to a hole in those bands, a
subsequent intraband transition may lead to a spin change. We also
compute the spin change along the W-U direction and find the results
are similar (not shown).

\subsection{Spin change along the  X-W  and U-X directions}

\begin{figure}
\centerline{\psfig{file=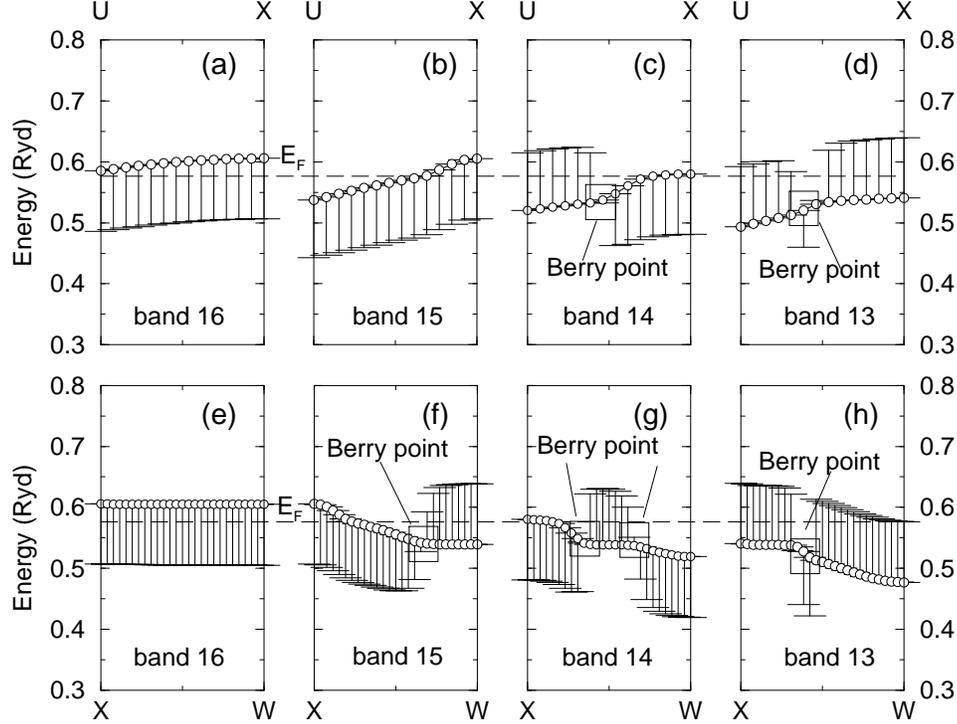,width=1\columnwidth,angle=0}}
\caption{Spin angular momentum change along the U-X and X-W
  directions. The longest bar corresponds to $\hbar/2$.  (a), (b),
  (c) and (d) refer to bands 16, 15, 14 and 13 along the U-X line,
  respectively. (e), (f), (g) and (h) denote bands from 16, 15, 14 and
  13 along the X-W direction.  Boxes highlight the spin Berry points.
  Before and after these points, the spin changes its direction.  }
\label{xuw}
\end{figure}

If we look at the Brillouin zone in Fig. \ref{fig1}(b), U, W and X are
on the same square.  The distance from the X point at (0,1,0) to U at
$(\frac{1}{4},1,\frac{1}{4})$ is closer than that to W at
$(\frac{1}{2},1,0)$. What is peculiar about these two directions is
that they both have a weakly dispersive unoccupied band, band 16, (see
Figs. \ref{xuw}(a) and \ref{xuw}(e)). This gives us another
opportunity to examine the spin change. We see that their spin angular
momentum points down and does not change much going from U to X and
from X to W. This finding is consistent with our finding above: The
less the band interaction is, the weaker the spin change becomes.
Figure \ref{xuw}(b) shows that the first band crosses the Fermi level
along the U-X direction, but there is no change in spin angular
momentum. This means that even if a laser field is applied along this
direction, the intraband transition won't bring in spin
change. However, the same band, band 15, along the X-W direction has a
clear spin change at a point slightly closer to the W point
(highlighted by a box in Fig. \ref{xuw}(f)), but it does not
cross the Fermi level at this box, so intraband transitions do not
play a role here. Nevertheless, this is very interesting.

\newcommand{\nnk}{n{\bf k}}

\newcommand{\nka}{n{\bf k}_1}
\newcommand{\nkb}{n{\bf k}_2}

Next we investigate band 14 along the U-X direction
(Fig. \ref{xuw}(c)), where we find a similar spin change in the middle
of the zone, highlighted again with a box. If we compare these two
boxes in Figs. \ref{xuw}(f) and \ref{xuw}(c), we notice that those
bands have a common feature where the band curvature changes at those
boxes, an infection point. These strange points appear in other bands
as well. Figure \ref{xuw}(g) has two such points, while band 13 along
both the U-X and X-W directions have one single point.  They always
appear at band infection points, which remind us of the Berry
phase.\cite{berry,zwanziger}

To be definitive, assume that $\psi_{\nnk}=e^{i{\bf k} \cdot {\bf r}}
\left (u^{\downarrow}_{\nnk}(\rr)\beta+u^{\uparrow}_{\nnk}(\rr)\alpha
\right )$, where $u^{\downarrow}_{\nnk}(\rr)$ and
$u^{\uparrow}_{\nnk}(\rr)$ are two periodic spatial functions in spin
minority and majority channels, and $\alpha$ and $\beta$ are the spin
up and spin down states.  We can rewrite the expectation value of spin
angular momentum as \be \la \psi_{n\bf k} |\hat{s}_z |\psi_{n {\bf
    k}}\ra =\frac{\hbar}{2} (\la u^{\uparrow}_{\nnk}
|u^{\uparrow}_{\nnk} \ra - \la u^{\downarrow}_{\nnk}
|u^{\downarrow}_{\nnk}\ra ). \ee Then the spin gradient $\nabla_{\bf
  k}s^z_{\nnk}$ that enters the spin momentum change in Eq. \ref{spin}
is \be \nabla_{\bf k}s^z_{\nnk}=\frac{\hbar}{2} \left (\la \nabla_{\bf
  k} u^{\uparrow}_{\nnk} |u^{\uparrow}_{\nnk}\ra + \la
u^{\uparrow}_{\nnk} |\nabla_{\bf k}u^{\uparrow}_{\nnk}\ra \right ) -
\left (\la \nabla_{\bf k} u^{\downarrow}_{\nnk}
|u^{\downarrow}_{\nnk}\ra + \la u^{\downarrow}_{\nnk} |\nabla_{\bf
  k}u^{\downarrow}_{\nnk}\ra \right ) \ee where the two terms are the
Berry potentials.\cite{vanderbilt} The spin gradient is just the
difference between the spin majority and minority Berry potentials, or
simply the spin Berry potential.
To highlight the importance of this Berry phase connection, in
Fig. \ref{xuw} we denote them as spin  Berry points. The significance of
these points is that before and after those points the spin changes
its direction, which is crucial to demagnetization. They occur around
those regions where two or more bands cross each other.  Each band 13 in
Figs. \ref{xuw}(d) and \ref{xuw}(h) has a single point.

\begin{figure}
\centerline{\psfig{file=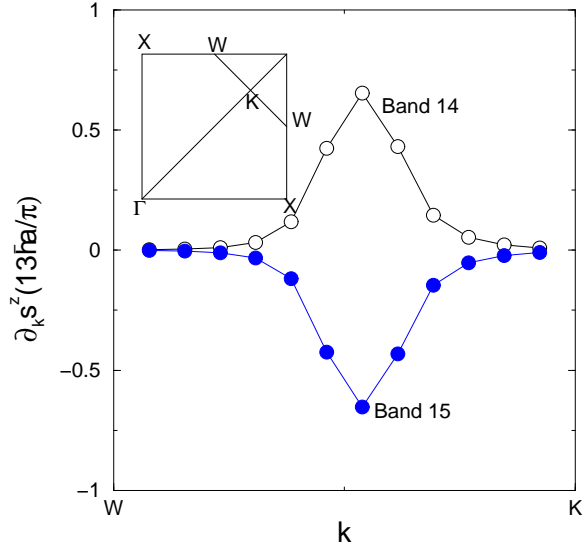,width=0.6\columnwidth,angle=0}}
\caption{ Spin angular momentum gradient $\nabla_{\bf k} s^z_{\nnk}$
  along the W-K line for bands 14 (open circles) and 15 (filled
  circles). $a$ is the lattice constant of fcc Ni. Inset: the square
  has four special points on the same plane. Note that the unit of the
  spin gradient is specific to our current {\bf k} mesh (see the text
  for an explanation). }
\label{berry}
\end{figure}

 Figure \ref{berry} shows the spin gradient along the W-K
 line.\footnote{We do not plot the spin gradient for U-X and X-W lines
   because band crossing makes it difficult to compute.}  The unit of
 the spin angular momentum gradient is $13\hbar a/\pi$.\footnote{This
   unit is calculated as follows. The spin angular momentum has a unit
   of $\hbar/2$. $\Delta k_{x(y)}=(2/104) \frac{2\pi}{a}$ since our W
   point is at (104,52,0)/104, and next point on the line is
   (102,54,0)/104 until we reach K at (78,78,0)/104, all in the unit
   of $\frac{2\pi}{a}$.}  The minimum $\Delta {\bf k}$ along the W-K
 line is $ (\frac{2}{104},\frac{2}{104},0)\frac{2\pi}{a}$, where $a$
 is the lattice constant of fcc Ni.  Now consider we have a laser
 pulse with electric field of 0.05 $\rm V/\AA$, which is commonly used
 experimentally.\cite{jpcm10} Figure \ref{berry} shows the gradient
 maximum is 0.653619 $\frac{13\hbar a}{\pi}$.  We plug this spin
 gradient and the laser electric field into Eq. \ref{spin}, and we
 find that for every fs, the spin changes by 0.724 $\hbar$.
 This spin change is huge.  However, for them to be operative,
 electrons must be able to move along the band from one {\bf k} point
 to another. This means that the state that an electron moves into
 must be vacated first. Bands 14 and 15, as they stand, are not those
 bands.  We will come back to this below with an experiment
 perspective.


\subsection{Spin change along the $\Gamma$-L and L-W directions}

\begin{figure}
\centerline{\psfig{file=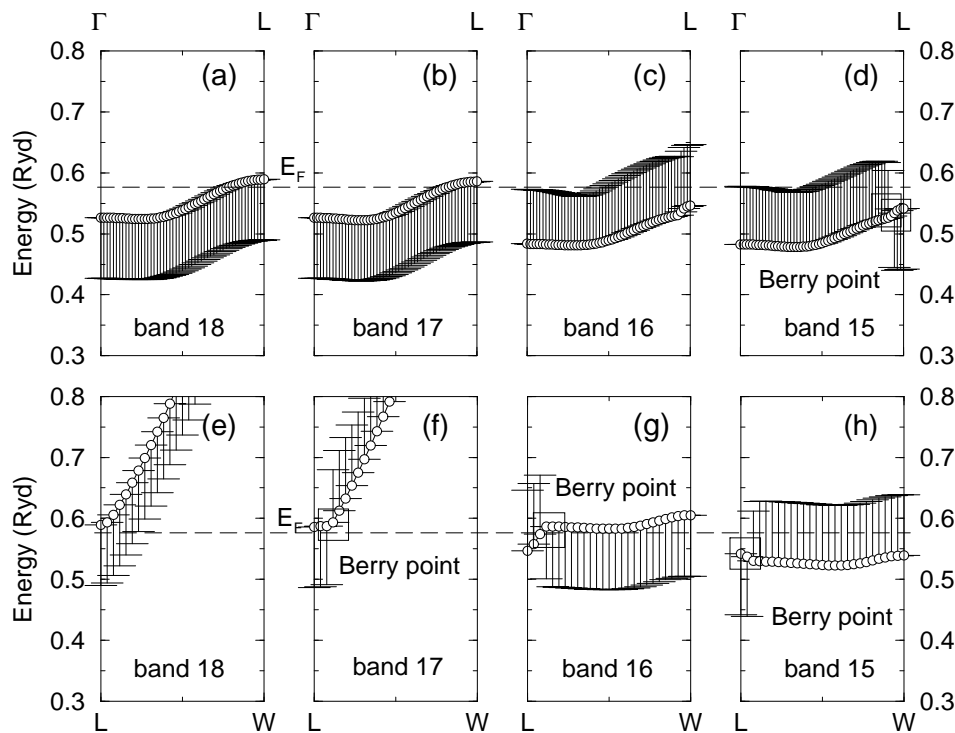,width=1\columnwidth,angle=0}}
\caption{Band dispersion overlapped with spin angular momentum denoted
  by the error bar, where the longest bar corresponds to $\hbar/2$.
  (a)-(d) shows the spin angular momentum along the $\Gamma$-L
  direction.  (e)-(h) shows the results along the L-W direction.  Spin
  Berry points are highlighted with boxes. The Fermi energy is denoted
  by horizontal dashed lines.  }
\label{l2w}
\end{figure}

Whether electron transport affects spin angular momentum depends on
whether intraband transitions pass through spin Berry points. Any
transition must rigorously obey the Pauli exclusion
principle. Consider two electrons in the same band $n$ at two
different points, ${\bf k}_1$ and ${\bf k}_2$ with occupations
$f_{\nka}$ and $f_{\nkb}$. For such an intraband transition to occur,
at least $f_{\nka}(1-f_{\nkb})$ or $f_{\nkb}(1-f_{\nka})$ or both
differ from zero. The spin angular momentum change is \be \Delta S^z=
f_{\nka}(1-f_{\nkb})(s^z_{\nkb}-s^z_{\nka})+
f_{\nkb}(1-f_{\nka})(s^z_{\nka}-s^z_{\nkb}), \ee in the limit that
$\nka$ and $\nkb$ are very close, \be \Delta S^z=\left[\nabla_{\bf
    k}f_{n{\bf k}} \cdot \nabla_{\bf k} s^z\right ] (\Delta k)^2, \ee
where $ \nabla_{\bf k} s^z$ is the spin gradient which connects with
the spin Berry potential. Physically, this means that to have a
nonzero spin change, the band occupation and spin must be changed
simultaneously. $\nabla_{\bf k}f_{n{\bf k}}$ is largest when the
state is close to the Fermi level. For this reason, we are mostly
interested in those bands close to the Fermi level.

Figures \ref{l2w}(a)-(d) shows that along $\Gamma$-L direction only
bands 18 and 17 cross the Fermi level and are partially occupied, but
there is no spin Berry point (see Figs. \ref{l2w}(a) and \ref{l2w}(b).
Bands 16 and 15 are completely filled (see Figs. \ref{l2w}(c) and
(d)), but a single spin Berry point is present for band 15. L is at
the center of the hexagon (Fig. \ref{fig1}(b)). Another high symmetry
line is the L-W line.  Figures \ref{l2w}(e)-(h) present two different
types of bands. Bands 18 and 17 are highly dispersive
(Figs. \ref{l2w}(e) and \ref{l2w}(f)), characteristic of $4sp$
bands. They barely touch the Fermi level. $4sp$ bands have fewer spin
Berry points, but their group velocities are larger.  A single spin
Berry point in band 17 (Fig. \ref{l2w}(f)) close to L is directly
related to band 16 (Fig. \ref{l2w}(g)). Because of band crossing, a
spin Berry point appears (Fig. \ref{l2w}(g)). Other than this, band 16
has no other Berry point. The situation in band 15 is similar to band
16 and is much less dispersive, characteristic of $3d$ bands. In
addition, it is fully filled as it is below the Fermi level. We also
compute the spin change along the $\Gamma$-K direction, where the
results are similar (not shown).

\subsection{Spin change along the $\Gamma$-X direction}

Spin change along the $\Gamma$-X direction is the most complicated and
also most interesting one among all the high symmetry lines
investigated. While we intend to investigate it first, we immediately
realize that it is hard to make sense of it in the beginning. Spin
Berry points are present among all the bands at different {\bf k}
points.

\begin{figure}
\centerline{\psfig{file=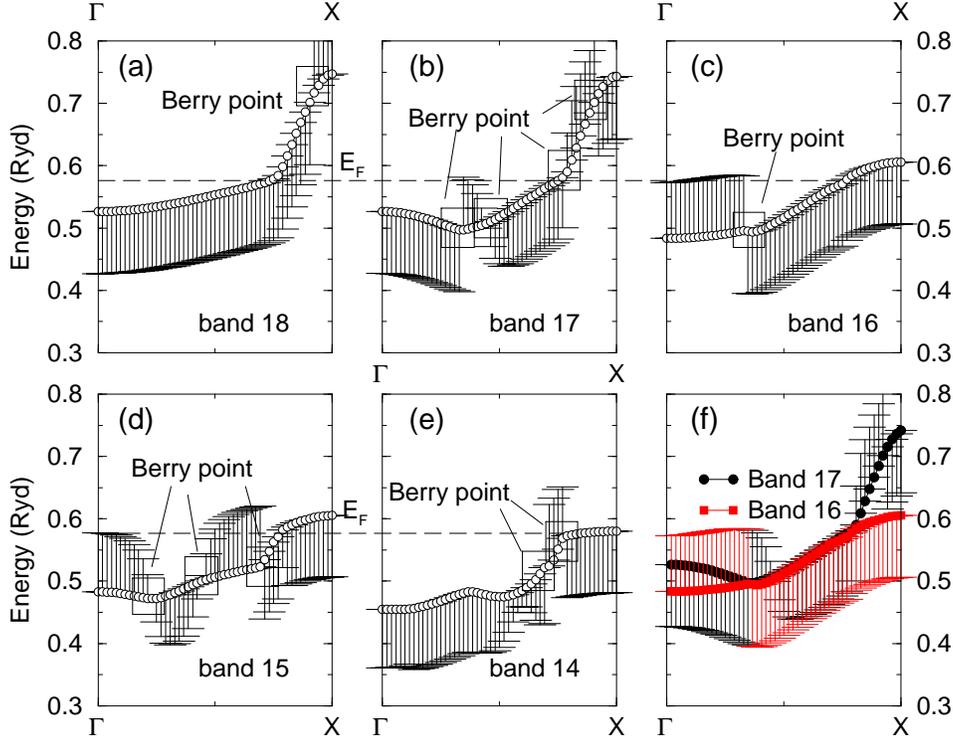,width=1\columnwidth,angle=0}}
\caption{ Spin structure along the $\Gamma$-X direction for five bands
  from band 18 to 14, where the longest bar corresponds to $\hbar/2$.
  (a) Band 18 cuts through the Fermi energy (horizontal dashed line)
  and has a spin Berry point close to X. (b) Band 17 has four spin
  Berry points, two below, one at and one above the Fermi level. (c)
  Band 16 has one point. (d) Band 15 has three. (e) Band 14 has
  two. (f) Comparison between bands 17 and band 16. Notice that the
  far left spin Berry point appears at the same location.  }
\label{gamma2x}
\end{figure}

Figure \ref{gamma2x}(a) shows that band 18 is partially occupied, and
close to the X point there is a spin Berry point above the Fermi
level. Band 17 in Fig. \ref{gamma2x}(b) has four spin Berry points,
the most among all the bands that we investigate. Two points are below
the Fermi level. The far left point is linked to the spin Berry point
on band 16 (Fig. \ref{gamma2x}(c)) at the same crystal momentum point.
Figure \ref{gamma2x}(f) compares them in the same graph, where the
first 17 {\bf k} points have spin pointing in opposite directions,
but this match ends very quickly since band 15 (Fig. \ref{gamma2x}(d))
starts to intrude in the same region of energy. These changes are
difficult to track if we do not highlight them with spin Berry
points. But they are not random. For instance, bands 15 and 14 also
interact in the same way. If we compare Figs. \ref{gamma2x}(d) and
(e), the far right spin Berry points are also at the same {\bf k}. We
believe these points are essential to laser-induced intraband
demagnetization and spin transport, which deserve extra scrutiny.

\section{Discussion and experimental connections}

Spin Berry points affect not only laser-induced demagnetization but
also spin transport across a magnetic sample, because they induce a
crucial spin flip when they are driven externally.  We need to link
them to actual experiments as how to probe them. But at first, we must
establish that these points are not random. Spin Berry points up to
now are presented band by band, so we can see them clearly as to how they
change across the Brillouin zone. However, this does not give the
entire picture.

\begin{figure}
\centerline{\psfig{file=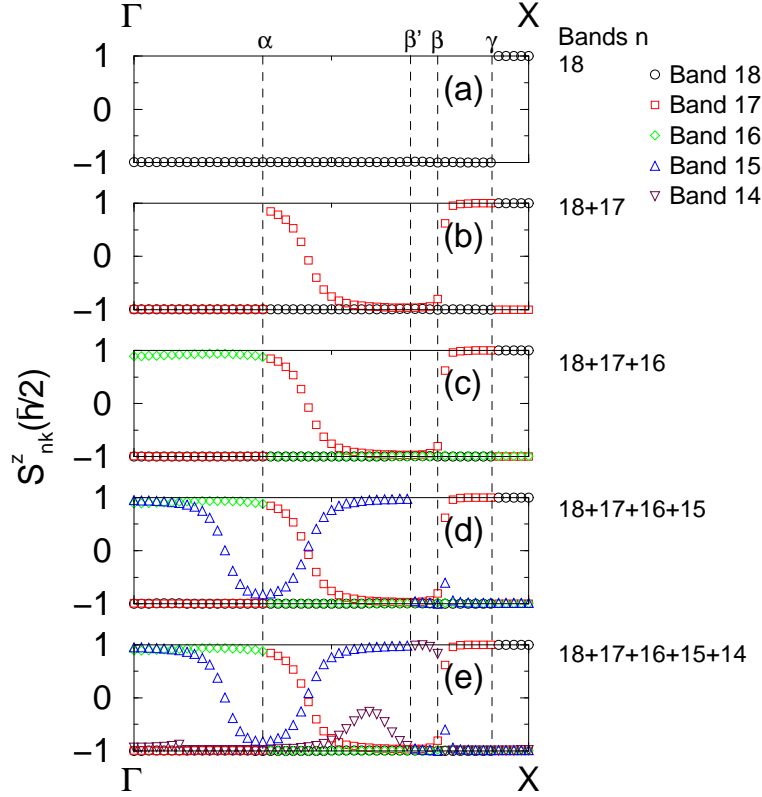,width=0.8\columnwidth,angle=0}}
\caption{ Establishing spin Berry lines by gradually including more
  bands. Three vertical dashed lines denote where spins should be
  connected.  (a) Only band 18 is present (circles). There is a sharp
  step jump at the $\gamma$ line. (b) Inclusion of band 17 softens the
  jump in spin by attaching more points at the $\gamma$ line (see boxes),
  but generates a new jump at the $\alpha$ line (boxes). (c) Band 16
  directly attaches the spin points at the $\alpha$ line (diamonds), so
  the jump does not appear anymore.  (d) Band 15 contributes a new
  spin curve (up triangles), but introduces a new jump at the $\beta$
  line. (e) Inclusion of band 14 removes the sharp jump at $\beta$
  line, in the same way as above.  Spin Berry points are linked among
  different bands, and are in fact spin Berry lines. There are five
  lines -- line 1: $\Gamma-\Diamond-\alpha-\Box-\gamma-\bigcirc-$X, line
  2: $\Gamma-\bigcirc-\alpha-\bigcirc-\gamma-\Box -$X, line 3:
  $\Gamma-\Box-\alpha-\Diamond-\gamma-\Diamond-$X; line 4:
  $\Gamma-\bigtriangleup-\beta'-\bigtriangledown-\beta-\bigtriangleup-$X,
  line 5:
  $\Gamma-\bigtriangledown-\beta'-\bigtriangleup-\beta-\bigtriangledown-$X.
}
\label{only}
\end{figure}

Figure \ref{only} shows a different picture along the $\Gamma$-X line,
where we increase one band at a time from Figs. \ref{only}(a) to
\ref{only}(e), where bands 18, 17, 16, 15 and 14 are denoted by
circles, boxes, diamonds, up triangles and down triangles,
respectively. Three vertical dashed lines $\alpha$, $\beta$, $\beta'$,
and $\gamma$ separate four spin Berry zones.  Figure \ref{only}(a)
shows that the spin angular momentum stays at $-\hbar/2$ (see circles)
but then when it hits the $\gamma$ line, it jumps from $-\hbar/2$ to
$\hbar/2$. It is difficult to understand why spin can have such a
radical change to flip spins suddenly. However, if we add band 17, it
immediately patches this spin jump at the $\gamma$ line with more
points (see boxes in Fig. \ref{only} (b)), so the jump is not steep
anymore.  This indicates that the spin angular momentum change is
shared by two bands. In other words, in the band structure, we
consider bands 18 and 17 are two different energy bands, but in the
spin space, some of them belong to the same spin
band. Microscopically, spin-orbit coupling does not permute with the
original Hamiltonian and creates two new sets of bands from two
spin-polarized bands, but the spin-part of the wavefunctions is still
correlated. This finding is interesting, so we decide to include more
bands.

We notice that Fig. \ref{only}(b) shows that band 17 introduces a new
spin jump at the $\alpha$ line.  Figure \ref{only}(c) shows that at the
$\alpha$ line the inclusion of band 16 complements the initial jump
left behind by band 17.  The diamonds before the $\alpha$ line smoothly
merge into boxes, and boxes further merge into circles at the $\gamma$
line, completing a single line, called, the spin Berry line.  The boxes
before the $\alpha$ line smoothly merge into diamonds along the way to the
X point, completing another spin Berry line. Band 15 in
Fig. \ref{only}(d) introduces a new line from $\Gamma$ to a point
close to the $\beta$ line (see up triangles), and then suddenly drops to
$-\hbar/2$. This sudden jump is removed by band 14 (down triangles in
Fig. \ref{only}(e)) at the same location. Band 14 has a peak between
the $\alpha$ and $\beta$ lines.  Along the $\Gamma$-X line, we can at
least identify five spin Berry lines -- line 1:
$\Gamma-\Diamond-\alpha-\Box-\gamma-\bigcirc-$X, line 2:
$\Gamma-\bigcirc-\alpha-\bigcirc-\gamma-\Box -$X, line 3:
$\Gamma-\Box-\alpha-\Diamond-\gamma-\Diamond-$X; line 4:
$\Gamma-\bigtriangleup-\beta'-\bigtriangledown-\beta-\bigtriangleup-$X,
line 5:
$\Gamma-\bigtriangledown-\beta'-\bigtriangleup-\beta-\bigtriangledown-$X.

It is important to make connections with experiments. Detection of
Berry's phase in BiTeI\cite{murakawa2013} and superconducting charge
pump\cite{mottonen2008} was reported. Our prediction does not require
those exotic phases.\cite{adak2020} There are two types of bands that
experimentally one can probe.  One type is where those bands directly
cross the Fermi level. They are partially occupied.  Suppose that one
can apply a voltage bias along the $\Gamma$-X direction for bands 14
and 17. Spins in this band can be switched from one direction to
another. If one employs a laser pulse, a similar effect can be
obtained as well. Optically, these are intraband transitions, where a
low energy photon, maybe in the infrared region, drives electrons
across the Fermi level.  This is probably the reason behind spin
flipping observed experimentally.\cite{turgut2013,eich2017} If this
happens between two materials with different spin polarization, spin
injection can be achieved.

The second type is related to interband transition. As seen in several
figures, there are fewer directions along which band states have spin
flipping. But in those bands that are away below the Fermi level,
multiple spin Berry points are noticed. These bands are occupied, so
without any excitation, electrons cannot move. A possibility arises that if
electrons are first excited to an excited state, then a subsequent
intraband transition may lead to a strong spin moment change in the
valence band. Time- and spin-resolved photoemission has the potential
to resolve spin change along different crystal momentum directions,
and finally identify those spin Berry points and Berry lines.

\section{Conclusion}

Different from many prior studies, we focus on how the spin angular
momentum changes with the crystal momentum for those bands that are
close to the Fermi surface.  We have employed fcc Ni as an example,
and we find that the spin dispersion with the crystal momentum has
abrupt jumps from $\pm \hbar/2$ to $\mp \hbar/2$, very unusual in the
first glance.  We notice those jumps only appear when two or more
bands are close by. These special crystal momentum points are called
spin Berry points, where the spin Berry potential for the spin
majority and spin minority bands is different. When we group several
bands together, we find these spin Berry points in fact form a smooth
spin Berry line, and there are many of these lines present along a
particular direction. Because these bands participate in intraband
transitions and spin transport, the information that we find here is
very important to both laser-induced ultrafast demagnetization and
spintronics.  We expect that time- and spin-resolved photoemission is
able to detect them.

This work was supported by the U.S. Department of Energy under
Contract No.  DE-FG02-06ER46304. We acknowledge part of the work as
done on Indiana State University's high-performance computers.  This
research used resources of the National Energy Research Scientific
Computing Center, which is supported by the Office of Science of the
U.S.  Department of Energy under Contract
No. DE-AC02-05CH11231. Calculations used resources of the Argonne
Leadership Computing Facility at Argonne National Laboratory, which is
supported by the Office of Science of the U.S. Department of Energy
under Contract No.  DE-AC02-06CH11357.

$^*$guo-ping.zhang@outlook.com.  https://orcid.org/0000-0002-1792-2701

\end{document}